\documentclass[aps,pre,twocolumn,showpacs]{revtex4}

\usepackage{graphicx}
\usepackage{amssymb,color,cancel}
\usepackage[normalem]{ulem}

\begin{document}

\title{Role of osmotic and hydrostatic pressures in bacteriophage genome ejection}

\author{Serge G. Lemay} 
\affiliation{MESA+ Institute for Nanotechnology, PO Box 217, 7500
  AE Enschede, The Netherlands}
\author{Debabrata Panja}
\email[]{D.Panja@uva.nl}
\affiliation{Institute of Physics, Universiteit van
  Amsterdam, Postbus 94485, 1090 GL Amsterdam, The Netherlands} 
\affiliation{Institute for Theoretical Physics, Universiteit Utrecht,
Leuvenlaan 4, 3584 CE Utrecht, The Netherlands}
\author{Ian J. Molineux}
\affiliation{Molecular Genetics and Microbiology, Institute for Cell and
  Molecular Biology, The University of Texas at Austin, TX 78712, USA}

\begin{abstract} 
  A critical step in the bacteriophage life cycle is genome ejection
  into host bacteria. The ejection process for double-stranded DNA
  phages has been studied thoroughly \textit{in vitro}, where after
  triggering with the cellular receptor the genome ejects into a
  buffer. The experimental data have been interpreted in terms of the
  decrease in free energy of the densely packed DNA associated with
  genome ejection. Here we detail a simple model of genome ejection in
  terms of the hydrostatic and osmotic pressures inside the phage, a
  bacterium, and a buffer solution/culture medium. We argue that the
  hydrodynamic flow associated with the water movement from the buffer
  solution into the phage capsid and further drainage into the
  bacterial cytoplasm, driven by the osmotic gradient between the
  bacterial cytoplasm and culture medium, provides an alternative
  mechanism for phage genome ejection \textit{in vivo}; the mechanism
  is perfectly consistent with phage genome ejection \textit{in
    vitro}.
\end{abstract}


\maketitle

\section{introduction}

Bacteriophages --- phages for short --- are viruses that infect
bacteria. In order to infect a host, the phage tail ($>95\%$ phage
virions contain tails) attaches to a bacterium, which triggers the
opening of the tail, and the genome (usually double-stranded DNA) is
delivered, in some cases together with protein molecules, into the
cytoplasm. Inside the capsid, the genome is packaged at near
crystalline density ($\sim$500 mg/ml for many double-stranded DNA
phages \cite{earn}), where it is in a state of high free energy due to
DNA self-interactions (causing high osmotic pressure within a mature
phage) and, to a lesser extent \cite{qiu}, the mechanical energy cost
incurred in bending the polymer into a compartment with a radius
comparable to DNA's persistence length. Energy for genome packaging is
provided by the bacterial cell in which the phage was assembled. From
a purely thermodynamic perspective, the stored free energy within the
capsid can provide the driving force for DNA ejection, with the DNA
acting essentially as a ``loaded spring'' that expands when opening of
the tail is triggered. (There may be separate gates at the junction of
the capsid and the tail tip, but we only consider the situation where
all gates are either open or closed.) The loaded spring concept has
been formalized in a continuum mechanics model
\cite{tzlil,purohit,inam}. Keeping in mind that the model applies only
to equilibrium states, and not to the dynamics or molecular mechanism
of the process, it successfully explains {\it in vitro\/} data for
phage $\lambda$ DNA ejection. Such experiments include the addition of
capsid-permeable DNA-condensing agents or capsid-impermeable osmolytes
to arrest ejection when the DNA remaining in the capsid reaches
equilibrium with the buffer \cite{alex1,gray1,alex2,wu} and
time-resolved light scattering \cite{lof}. In addition, fluorescence
microscopy experiments following {\it in vitro\/} $\lambda$ DNA
ejection in real time at the single-molecule level show that the
genome is ejected in a single step at an ion-dependent rate, but one
that is not a simple function of the amount of DNA remaining in the
capsid \cite{wu,gray2}.

Experimental data on osmotic suppression of DNA ejection of phage SPP1
\cite{jose} are consistent with the continuum mechanics model, but
those on T5 are less so. The extent of T5 DNA ejected after tail
triggering is not proportional to low external osmotic pressure and
different populations co-exist at equilibrium \cite{lef1}. Further,
DNA remaining in individual capsids was shown to undergo several
phase transitions as ejection proceeds \cite{lef2}, and there are
distinct pauses where DNA exit from individual particles is
temporarily halted \cite{mang,chia}. These pauses do not correspond to
the natural nicks in the T5 genome, which are also not required for
pausing, and each re-starting event requires overcoming the same
activation energy \cite{chia,defrutos,rasp}. It is not known what
causes pausing during T5 DNA ejection.

It is natural to imagine that the loaded spring concept extrapolates
directly to DNA ejection {\it in vivo}, where the ejection medium
consists of the bacterial cytoplasm instead of a buffer solution. It
has, however, recently been argued that many features of phage DNA
ejection {\it in vivo\/} are incompatible with this simple concept
\cite{mol,gray3,panja}. These include the problems that the osmotic
pressure inside growing bacteria should prevent complete ejection of
the genome, the efficiency of phage infection being reduced when
bacterial turgor is low (the stationary phase of bacterial growth),
ejection of proteins before (and also likely after) DNA ejection, and
that some phages do not contain DNA packaged to high
density. Furthermore, several phage types are known to eject their DNA
slowly into the bacterial cytoplasm in two or more distinct
phases. These observations suggest mechanism(s) other than the loaded
spring may contribute to DNA ejection {\it in vivo}.

\begin{figure} 
\begin{center}
\includegraphics[width=\linewidth]{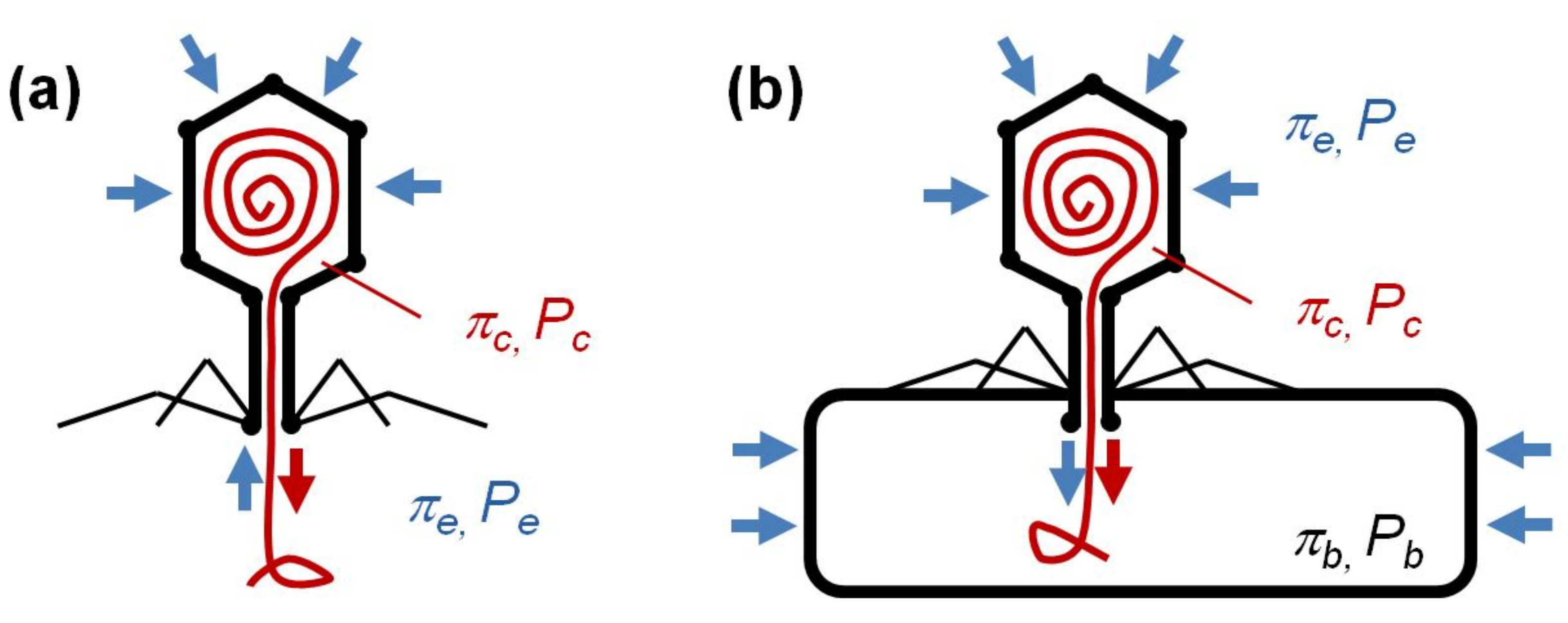}
\end{center}
\label{cartoon} \caption{(a) \emph{In vitro} configuration involves
  capsid and environment compartments, each with its own osmotic
  ($\pi$) and hydrostatic ($P$) pressure. As DNA is ejected (red
  arrow), its volume is replaced by water permeating through the
  capsid and flowing in through the tail (blue arrows). (b) \emph{In
    vivo} configuration with a third compartment, the interior of the
  bacterium. As the bacterium grows, it has a high internal osmotic
  pressure, $\pi_b$, and pulls in water. The resulting water flow
  through the tail can have the opposite direction as in (a) and can
  drive DNA ejection, independently of the osmotic pressure inside the
  capsid.}
\end{figure}

{\it In vitro\/} ejection into a buffer is a two compartment process
(capsid, bacterium), with distinct osmotic and hydrostatic pressures
in each compartment. {\it In vitro}, the net thermodynamic potential
gradient for water movement up the tail-tube, which directly connects
the two compartments, is positive, so water tends to flow up the
tail-tube (see later). Ejection into a bacterium ({\it in vivo\/}),
however, differs in subtle ways from ejection into a buffer. In
particular, ejection {\it in vivo\/} involves three distinct
compartments (bacterium, capsid, environment) instead of two {\it in
  vitro\/}. This is sketched in Fig. \ref{cartoon}. Each compartment,
again, has distinct osmotic and hydrostatic pressures, with
ramifications for the transport of the genome out of the phage.

Growing bacteria continuously draw in water from the environment. For
this to occur, a pressure imbalance (osmotic and/or hydrostatic) must
exist between the bacterium and the environment. We argue that the
capsid takes osmotic and hydrostatic pressure values that are
intermediate to those of the environment and the cytoplasm, leading to
water always flowing down the tail-tube. While the rate of water entry 
into the cytoplasm through the tail tube is at least three orders of 
magnitude smaller than that through the cytoplasmic membrane, it can 
exert significant local forces on DNA inside the tail tube. Further, 
the pressure imbalance represents a separate driving force for genome 
ejection, independent of the physics of the loaded spring model. The
hydrodynamic model of genome ejection \cite{mol,gray3,panja} is
consistent with many of the {\it in vivo} observations that are not
readily explained by the loaded spring concept \cite{panja}. We
further argue below that the model has ramifications  for the
interpretation of {\it in vitro\/} data, in particular the dynamics of
ejection.

To date, the hydrodynamic model has remained relatively qualitative
\cite{gray3,panja}. The purpose of the present article is to present a
simple explicit formulation of the model that retains its key physical
ingredients. Particular emphasis is given to the fact that the
hydrostatic pressure inside the viral capsid is free to adjust in
response to DNA ejection, a factor that has been largely ignored. We
show that such an analysis can assist the interpretation of existing
experiments, as well as guide the creation of artificial model systems
in which the physics of genome ejection can be further elucidated.

\section{Description of the model\label{sec2}}

In general, we consider three compartments, namely, the inside of the
capsid, the bacterial cytoplasm and the surrounding culture medium
(environment).  We denote the osmotic and the hydrostatic pressures of
these compartments by $\{\pi_c,P_c\}$, $\{\pi_b,P_b\}$ and
$\{\pi_e,P_e\}$, respectively. The variables $\pi_e$ and $P_e$ are
treated as constants that characterize the environment; $\pi_b$ and
$P_b$ similarly define the state of the bacterial cytoplasm,
$\pi_c$ is a monotonically decreasing function of the amount of DNA
remaining in the capsid, and $P_c$ is free to adjust to the dynamics
of DNA ejection.  The thermodynamic potential for water movement from
compartment $2$ to compartment $1$ is simply given by
\begin{equation}
  \phi_{12}=\underbrace{(\pi_1-P_1)}_{\phi_1}-\underbrace{(\pi_2-P_2)}_{\phi_2}=-\phi_{21}.
\label{e1}
\end{equation}

Assuming for simplicity that both water and the capsid are
incompressible, during the ejection process the continuity equation
for water movement implies that the rate at which DNA volume exits the
capsid equals the volume rate $W(t)$ of the net water entry into the
capsid at all times. Water entry into (or exit out of) the capsid can
occur via two distinct paths, namely, by permeating through the capsid
or by flowing through the tail tube.  Having noted that the available
cross-section of the tail-tube is much smaller than the capsid
cross-section, we assume that permeation through the capsid is the
dominant contribution to $ W(t)$. 
%For {\it in vivo\/} ejection it can
%be argued that the water entry into the cytoplasm through the
%cytoplasmic membrane is at least three orders of magnitude larger than
%that through the tail-tube. 
This simplification can easily be relaxed,
but only at the cost of complicating the analytical form of the final
results without yielding additional insights into the underlying
physics. We further assume that (i) ejection is sufficiently slow that
all compartments (bacterium, capsid and environment) remains
well-stirred (or equivalently, the water is fully equilibrated) in all
compartments at all stages of ejection, i.e., the osmotic and
hydrostatic pressures are uniquely, and uniformly, defined in all
compartments, and (ii) water entry into the capsid from the
environment is in the linear response regime. Under these conditions,
\begin{equation}
  W(t) =
A_{\text{DNA}}V_{\text{DNA}}(t) \approx \sigma_c\, \phi_{ce}(t).
\label{e3}
\end{equation}
Here $A_{\text{DNA}}$ is the cross-sectional area of the DNA,
$V_{\text{DNA}}(t)$ is the speed of DNA ejection, and $\sigma_c$ is
the capsid's permeability to water.

Having neglected the rate of water flow through the tail tube in
comparison to the rate of water entry into the capsid from the
environment through the capsid shell, the dynamics of genome ejection
both {\it in vitro\/} and {\it in vivo\/} are qualitatively the same;
namely that water osmosis into the capsid from the environment is
accompanied by DNA ejection from the capsid. The physical force
responsible for DNA ejection is the hydrostatic pressure gradient
acting on the DNA across the tail tube, and the corresponding work
done in ejecting the DNA is derived from the free energy made
available as water crosses the capsid shell [from the environment
  (with osmotic pressure $\pi_e$) into the capsid (with osmotic
  pressure $\pi_c$, where $\pi_c>\pi_e$)].
\begin{figure}[h]
\begin{center}
\includegraphics[width=0.5\linewidth]{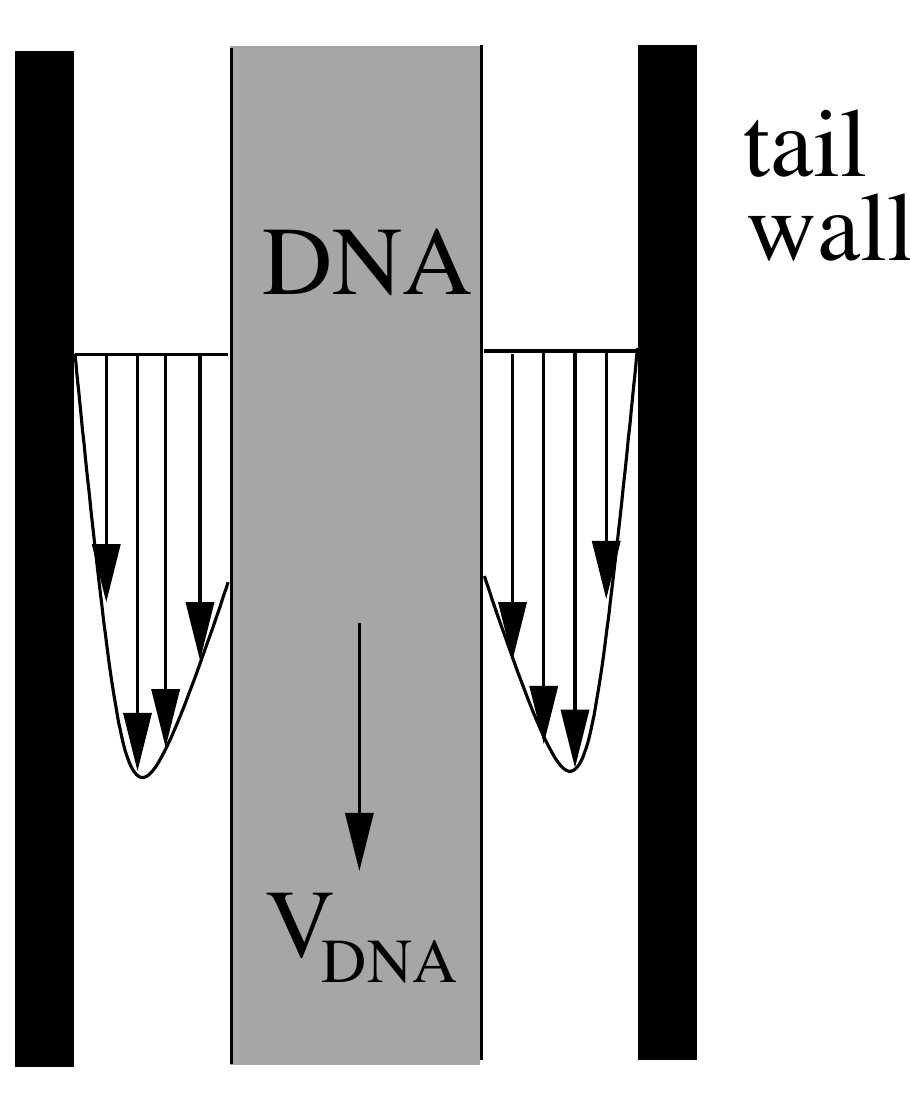}
\end{center}
\caption{The flow velocity pattern of water through the gap between
the DNA and the inner wall of the tail-tube, shown via a longitudinal
cross-section.\label{invivotail}}
\end{figure}

Although we neglect the contribution of the hydrodynamic flow of water
in the tail for the purpose of determining the total flux of water
into the capsid, flow in the tail {\it does\/} influence conditions
locally. Specifically, any flow of water in the tail, $v_w$, exerts a
force on DNA, also located in the tail, via hydrodynamic drag.  In
order to estimate this force, we represent the DNA as a cylinder of
radius $R_0$ (i.e., $A_{\text{DNA}}=\pi R_0^2$), and let the axis of
the moving DNA coincide with the axis of the tail tube at all times
during ejection. We denote the inner radius of the tail tube by
$R$. We further apply no-slip boundary conditions at the surface of
the DNA and at the inner surface of the tail tube.  Under this
simplified geometry, the velocity profile for water in the tail tube,
$q(r)$, exhibits cylindrical symmetry and is only a function of $r$,
the radial distance measured from the axis of the DNA, and $q(r)$ can
be easily determined by solving the Stokes equation, $\eta
r^{-1}\partial_r(r \partial_r q(r)) = \phi/L$ (with no-slip boundary
conditions). Here $\eta$ is the (dynamic) viscosity of water, $L$ is
the length of the tail-tube and $\phi$ represents $\phi_{ce}$ or
$\phi_{cb}$ for \textit{in-vitro} and \textit{in vivo},
respectively.  The resulting flow velocity pattern of water through
the gap between the DNA and the tail-tube, shown schematically in
Fig. \ref{invivotail}. The magnitude of the corresponding hydrodynamic
drag exerted by the fluid on the outgoing DNA, $F_{h} = 2\pi R_0L
\eta(\partial q /\partial r)_{r=R_0}$, is
\begin{equation}
F_{h} = \underbrace{\frac{\pi}{2} 
\left[ \frac{R^2-R_0^2}{\ln (R/R_0)} - 2R_0^2 \right]}_{\alpha_1}\,\phi
+ \underbrace{\frac{2\pi L\eta}{\ln (R/R_0)}}_{\alpha_2} V_{\text{DNA}}.
\label{f_hydro}
\end{equation}
Note that with $\phi$ being a combination of fixed parameters and the 
freely adjustable capsid pressure, the rhs of
Eq. (\ref{f_hydro}) contains two unknowns: the DNA velocity,
$V_{\text{DNA}}$, and the hydrostatic pressure of the capsid,
$P_c$. In order to uniquely determine these quantities, we treat the
\textit{in vitro} and \textit{in vivo} cases separately.

\section{Hydrodynamic model for \emph{in vitro} ejection}

In the {\it in vitro\/} experiments the phage is immersed in a buffer
solution, creating a two-compartment configuration: one compartment is
the inside of the capsid, while the other is the environment (around
the capsid).  Before the opening of the tail is triggered, there is no
net movement of water between these two compartments, consequently
$\phi_{ce}\equiv0$ and \cite{panja}
\begin{equation}
P_c=P_e+(\pi_c-\pi_e).
\label{e2}
\end{equation}
This is analogous to the classic case of the van 't Hoff water columns.

Following triggering of the tail, DNA begins to exit the capsid while
water simultaneously enters the capsid from the environment.  In order
to obtain an explicit expression for $V_{\text{DNA}}(t)$, it is
necessary to specify the relationship between $V_{\text{DNA}}(t)$ and
the current state of the system. In principle, this requires knowledge
of the free energy stored in the configuration of the DNA confined
inside the capsid. This free energy is a sum of the interaction energy
of the capsid content (DNA, water, ions and possibly other protein
molecules), and the mechanical energy stored in the DNA due to its
packaging in tight circles. The ``mechanical'' energy is however an
order of magnitude smaller than the contribution from interactions
\cite{qiu}; in the spirit of constructing the simplest model that
captures the role of hydrodynamics, we therefore neglect it here. With
this simplification, the force acting on the partially ejected DNA
that experiences a hydrostatic pressure difference $P_{ce}(t)=P_c(t)-P_e$ across
the tail-tube is given by $A_{\text{DNA}} P_{ce}(t)$.

Thus, having denoted the effective mass of the mobile DNA by $m$
(where it is reasonable to assume that $m$ is of the order of the mass
of the DNA cylinder within the tail tube) and combining
Eqs. (\ref{e1}), (\ref{e3}) and (\ref{f_hydro}), we have
\begin{eqnarray}
m\dot V_{\text{DNA}}(t)&=&P_{ce}(t)A_{\text{DNA}}-F_h(t),\quad\mbox{i.e.,}\nonumber\\
\frac{m\dot V_{\text{DNA}}(t)}{A_{\text{DNA}}}\!\!\!&=&\!\!\!\pi_{ce}(t)\!-\!\!\underbrace{\left[\frac{\alpha_1\!+\!A_{\text{DNA}}}{\sigma_c}\!+\!\frac{\alpha_2}{A_{\text{DNA}}}\right]}_\beta\!\! V_{\text{DNA}}(t)\!,
\label{vdotdna}
\end{eqnarray}
and correspondingly, 
\begin{equation}
P_c(t) = P_e + \pi_{ce}(t)-V_{\text{DNA}}(t)A_{\text{DNA}}/\sigma_c.
\label{pc}
\end{equation}
Equation (\ref{vdotdna}) is easily solved as an integral equation for
$V_{\text{DNA}}(t)$, in which $m/(\beta A_{\text{DNA}})$ sets the
(transient) time-scale beyond which the inertial term $m\dot
V_{\text{DNA}}(t)$ term is not relevant. The transient time-scale for
typical tail-tube diameters and lengths is estimated to be in the
range of microseconds. Given that the time-scales for {\it in vitro\/}
ejection is in the range of seconds \cite{gray2}, we neglect the
intertial term, which leads us to
\begin{eqnarray}
  V_{\text{DNA}}(t)\!=\!\frac{\pi_{ce}(t)}{\beta},\,\mbox{and}\,
  P_c(t)\!=\!P_e\!+\!\left[1\!-\!\frac{A_{\text{DNA}}}{\beta\sigma_c}\right]\pi_{ce}(t).
\label{approxsol}
\end{eqnarray}

For the hydrostatic pressure $P_c$ within the capsid, the result
(\ref{approxsol}) points to two distinct possible regimes. (a) For
high capsid permeability to water ($\beta\sigma_c\gg A_{\text{DNA}}$)
the hydrostatic pressure gradient $P_c(t)-P_e$ simply follows the
osmotic gradient $\pi_c(t)-\pi_e$ in a manner analogous to
Eq.~(\ref{e2}). (b) In the opposite limit, i.e., for low capsid
conductivity, the hydrostatic pressure gradient, $P_c(t)-P_e$,
decreases abruptly when the DNA starts to eject following the
triggering of the tail. The latter is the scenario considered for {\it
  in vitro\/} DNA ejection in Ref. \cite{panja}.

Which limit, or where between the limits one should observe, is not
{\it a priori\/} obvious. Given that the choice depends on the
relative ease of water and DNA transport between the capsid and the
environment, the choice likely depends on the phage studied. Some
phages are resistant to severe osmotic downshifts; for example, T7
infects normally when a suspension in $\sim5$M CsCl is added directly
to cells. Conversely, the classic Kleinschmidt picture of T2 DNA
emanating from a ruptured capsid is a result of rapid dilution of a
phage suspension in 2M ammonium acetate \cite{klein}. The resistance
to osmotic downshift should correlate well to the mobility of water
through the capsid membrane. In that case we would expect T7 to
conform more closely to limit (a) than T2, which would be closer to
limit (b).

The final important point to note is that when all motions of the DNA
and water cease, i.e., at the end of ejection (partial or complete as
it may be), then Eq. (\ref{approxsol}) indicates that both
osmotic and hydrostatic pressure gradients between the environment and
the capsid become zero. These situations correspond to the osmotic
suppression experiments \cite{alex1,gray1,alex2,wu,jose,lef1};
complete suppression of ejection yields an estimate of the osmotic
pressures in a mature phage particle: $\sim25$ atm for $\lambda$,
$\sim47$ atm for SPP1, and $\sim16$ atm for the deletion mutant
T5st(0).

\section{Hydrodynamic model for \emph{In vivo} ejection}

For {\it in vivo\/} situations, it is important to appreciate that
water uptake by the cytoplasm from the culture medium is very tightly
regulated in a growing bacterium, which maintains an active osmotic
gradient between the environment and the cytoplasm.  A corresponding
hydrostatic pressure gradient (turgor) helps the bacterium enlarge
\cite{koch}. For instance, the {\it E. coli\/} cytoplasm has a
positive osmotic pressure of several atm above the environment (under
various growth conditions the pressure has been estimated to vary
between 2 and 15 atm, with 3.5-5 atm being commonly accepted values
\cite{stock,koch1}). This osmotic pressure gradient is
counter-balanced by a hydrostatic pressure differential (turgor)
\cite{koch} that enables the cell to enlarge during growth
\cite{koch2}. Gram-positive cells, such as {\it Bacillus subtilis},
the host for phage SPP1, have much higher turgor: $\sim19$ atm
\cite{what}. When the cell membrane is breached by an infecting phage,
a new conduit opens up for water to enter from the environment into
the cell cytoplasm by passing through the phage capsid. For the
following analysis, we treat the cytoplasm as a uniform, well-stirred
medium for which the osmotic and the hydrostatic pressures $\pi_b$ and
$P_b$ remain constant during ejection. This is reasonable as bacteria
respond to osmotic perturbations within milliseconds. 

This situation is analogous to the {\it in vitro\/} case, with two
important modifications: First, the thermodynamic potential difference
for water transport between the environment (culture medium) and the
bacterial cytoplasm, $\phi_{be}$, which, in the light of the above
paragraph, \textit{is a strictly positive quantity\/} for a growing
bacterium, is given by
\begin{eqnarray}
\phi_{be}=\phi_{bc}+\phi_{ce}.
\label{phieb}
\end{eqnarray}
Secondly, the hydrodynamics in the tail tube is driven by the
themodynamic potential difference $\phi_{cb} (=-\phi_{bc})$ for water
transport from the capsid into the bacterial cytoplasm, i.e.,
\begin{eqnarray}
  F_h(t)=\alpha_1\phi_{cb}(t)+\alpha_2V_{\text{DNA}}.
\label{foftinvivo}
\end{eqnarray}
We then use the continuity of water flow across the capsid shell
(\ref{e3}) to write
\begin{eqnarray}
m\dot V_{\text{DNA}}(t)&=&P_{cb}(t)A_{\text{DNA}}-F_h(t),\quad\mbox{i.e.,}\nonumber\\
\frac{m\dot
V_{\text{DNA}}(t)}{A_{\text{DNA}}}&=&\pi_{cb}(t)+\underbrace{\frac{\alpha_1+A_{\text{DNA}}}{A_{\text{DNA}}}}_\gamma\phi_{be}\nonumber\\&&-\underbrace{\left[\frac{\alpha_1\!+\!A_{\text{DNA}}}{\sigma_c}\!+\!\frac{\alpha_2}{A_{\text{DNA}}}\right]}_\beta V_{\text{DNA}}(t).
\label{vdotdna2}
\end{eqnarray}
Once again, we neglect the inertial term [i.e., the transient time
$m/(\beta A_{\text{DNA}})$], and obtain the expressions for $V_{\text{DNA}}$ and
$P_c$ as
\begin{eqnarray}
  V_{\text{DNA}}(t)&=&\frac{\pi_{cb}(t)+\gamma\phi_{be}}{\beta},\,\mbox{and}\,\nonumber\\
  P_c(t)\!\!&=&\!\!P_e\!+\!\left[1\!-\!\frac{A_{\text{DNA}}}{\beta\sigma_c}\right]\pi_{ce}(t)\!-\!\frac{\gamma
    A_{\text{DNA}}}{\beta\sigma_c}\phi_{be}.
\label{approxsolinvivo}
\end{eqnarray}
Equation (\ref{approxsolinvivo}) shows that the rate of DNA ejection
is enhanced, irrespective of the osmotic pressure difference between
the capsid and the bacterium; the driving force behind this
enhancement being the hydrodynamic drag of water on the DNA in the
tail tube, while the source of free energy being the osmotic imbalance
between the bacterium and its environment \cite{mol,gray3,panja}. As
pointed out in Ref. \cite{panja}, $\phi_{be}$ aiding the DNA ejection
{\it in vivo\/} is consistent with infection phenomenology. No phage
can complete genome ejection without the $\phi_{be}$ term in
Eq. (\ref{approxsolinvivo}) \cite{nrm}. It is worthwhile to note in
this context that the DNA genome of phage P2 and its  relatives is
packaged at a lower density than many other phages, and a P4 genome
has been packaged in a P2 capsid to yield an infective particle {\it
  in vivo\/} \cite{pruss}. The genome length of P4 is only one-third
that of P2 and the internal osmotic pressure is less than is necessary
to even initiate DNA release into the cytoplasm with only $\pi_{cb}$
at its disposal. Note also that upon setting $\phi_{be}=0$, we recover
the {\it in vitro\/} results as of course we should.

\section{Conclusions}

We have introduced a simple model of genome ejection in terms of the
hydrostatic and osmotic pressures inside a phage, a bacterium, and the
surrounding medium. This model shows explicitly that the osmotic
balance between a bacterium and its environment provides a second
source of free energy for driving genome ejection that is independent
of the free energy stored in the configuration of the DNA in the
capsid. The model also demonstrates that water transport into the
capsid and hydrodynamic drag on the DNA in the tail pipe can
profoundly influence the rate of genome ejection, both {\it in vitro}
and {\it in vivo}. Further refinements to the hydrodynamic model in
the simple form described here can be envisioned, for example
accounting for a finite capsid elasticity or for the mechanical energy
stored in the DNA, and this may prove necessary for a quantitative
comparison with experiments. However, such refinements are not
expected to significantly change the physical picture described here.

\begin{acknowledgments}
  We thank C. M. Knobler and W. M. Gelbart for insightful
  discussions. SGL acknowledges financial support from the European
  Research Council (ERC) and the Netherlands Organization for
  Scientific Research (NWO).
\end{acknowledgments}


\begin{thebibliography}{phages}

\bibitem{earn} W. C. Earnshaw and S. R. Casjens, Cell {\bf 21}, 319
  (1980).

\bibitem{qiu} X. Qiu {\it et al.}, Phys. Rev. Lett. {\bf 106}, 028102
  (2011). 

\bibitem{tzlil} S. Tzlil, J. T. Kindt, W. M. Gelbart and A. Ben-Shaul,
  Biophys J. {\bf 84}, 1616 (2003).

\bibitem{purohit} P. K. Purohit {\it et al.}, Biophys J. {\bf 88}, 851
  (2005).

\bibitem{inam} M. M. Inamdar, W. M. Gelbart and R. Phillips, Biophys.
  J. {\bf 91} 411 (2006).

\bibitem{alex1} A. Evilevitch {\it et al.}, Proc. Nat. Acad. Sci. USA
  {\bf 100}, 9292 (2003).

\bibitem{gray1} P. Grayson {\it et al.}, Virology {\bf 348}, 430
(2006).

\bibitem{alex2} A. Evilevitch {\it et al.}, Biophys. J. {\bf 94} 1110
  (2008).

\bibitem{wu} D. Wu, D. van Valen, Q. Hu and R. Phillips, Biophys. J.
  {\bf 99}, 1101 (2010).

\bibitem{lof} D. L\"of, K. Schill\'en, B. J\"onsson and A. Evilevitch,
  J. Mol. Biol. {\bf 368}, 55 (2007).

\bibitem{gray2} P. Grayson, L. Han, T. Winther and R. Phillips, Proc.
  Nat. Acad. Sci. USA {\bf 104}, 14652 (2007).

\bibitem{jose} C. S\~ao-Jos\'e {\it et al.}, J. Mol. Biol. {\bf 374},
  346 (2007).

\bibitem{lef1} A. Leforestier {\it et al.}, J. Mol. Biol. {\bf 384},
  730 (2008).

\bibitem{lef2} A. Leforestier and F. Livolant, J. Mol. Biol. {\bf
    396}, 384 (2010).
 
\bibitem{mang} S. Mangenot, M. Hochrein, J. R\"adler and L.
  Leterllier, Curr. Biol. {\bf 15}, 430 (2005).

\bibitem{chia} N. Chiaruttini {\it et al.}, Biophys J. {\bf 99} 447
  (2010).

\bibitem{defrutos} M. de Frutos, L. Letellier and E. Raspaud, Biophys.
  J. {\bf 88}, 1364 (2005).

\bibitem{rasp} E. Raspaud {\it et al.}, Biophys. J. {\bf 93}, 3999
  (2007).

\bibitem{mol} I. J. Molineux, Virology {\bf 344}, 221 (2006).

\bibitem{gray3} P. Grayson and I. J. Molineux , Curr. Op. in
  Microbiol. {\bf 10}, 401 (2007).

\bibitem{panja} D. Panja and I. J. Molineux, Phys. Biol. {\bf 7},
  045006 (2010).

\bibitem{klein} A. K. Kleinschmidt, D. Lang D, D. Jacherts and
  R. K. Zahn, Biochim. Biophys. Acta {\bf 1000} 41-48 (1989). This is
  a reprint of the original article in German [A. K. Kleinschmidt,
  D. Lang D, D. Jacherts and R. K. Zahn, Biochim. Biophys. Acta {\bf
    61} 857 (1962)].

\bibitem{koch} A. L. Koch, Crit. Rev. Microbiol. {\bf 24}, 23 (1998). 

\bibitem{stock} J. B. Stock, B. Rauch and S. Roseman, J. Biol.
  Chem. {\bf 252}, 7850 (1977).

\bibitem{koch1} A. L. Koch, Adv. Microbiol. Physiol. {\bf 24}, 301
  (1983).

\bibitem{koch2} A. L. Koch, J Bacteriol. {\bf 159} 919 (1984).

\bibitem{what} A. M. Whatmore and R. H. Reed, J. Gen. Microbiol. {\bf
    136} 2521 (1990).

\bibitem{nrm} I. J. Molineux and D. Panja, Nat. Rev. Microbiol. {\bf 11},
194 (2013). 

\bibitem{pruss} G. Pruss, R. N. Goldstein and R. Calendar,
  Proc. Natl. Acad. Sci. (USA) {\bf 71}, 2367 (1974). 

\end{thebibliography}
\end{document}